# Enhanced 5G/B5G Network Planning/Optimization deploying RIS in Urban/Outdoor Scenarios

Valdemar R. Farré-Guijarro*, Juan C. Estrada-Jiménez†, José D. Vega Sánchez‡, Juan A. Vasquez-Peralvo§, Symeon Chatzinotas§

*Departamento de Electrónica, Telecomunicaciones y Redes de Información Escuela Politécnica Nacional (EPN), Quito
†IT for Innovative Services Department, Luxembourg Institute of Science and Technology, Luxembourg
‡Colegio de Ciencias e Ingenierías "El Politécnico",Universidad San Francisco de Quito (USFQ), Diego de Robles s/n, Quito 170157
§Interdisciplinary Centre for Security Reliability and Trust, University of Luxembourg, Luxembourg
`valdemar.farre@epn.edu.ec, juan.estrada-jimenez@list.lu,`
`dvega@usfq.edu.ec, juan.vasquez@uni.lu, symeon.chatzinotas@uni.lu`

*Abstract*—In recent years, the fifth-generation (5G) mobile network has been developed worldwide to remarkably improve network performance and spectral efficiency. Very recently, reconfigurable intelligent surfaces (RISs) technology has emerged as an innovative solution for controlling the propagation medium of the forthcoming sixth-generation (6G) networks. Specifically, RIS takes advantage of the reflected rays on the propagation environment to redirect them to a desired target, improving wireless coverage. To further improve RIS performance, an interesting technique called synchronized transmission with advanced reconfigurable surfaces (STARS) has appeared to allow simultaneous transmission and reflection of intelligent omni-surfaces. With that in mind, this paper introduces an enhanced strategy for the network planning of 5G and beyond (B5G) mobile networks in dense urban scenarios focused on the city of Quito-Ecuador. The efficacy of RIS and its cutting-edge STARS concept is emphasized, providing useful insights into coverage, quality, and throughput results. In particular, this work considers the $3.5/28$ GHz frequency bands, optimizing the radio network and anticipating their applicability in B5G networks. Finally, simulation results are also shown, which allow the identification of the benefits of STAR RIS in terms of coverage, signal quality, and data performance.

## I. INTRODUCTION

The rapid evolution of wireless communication networks, particularly in the dynamic landscape of the fifth-generation (5G) and beyond 5G (B5G) networks, marks a significant milestone in mobile communications. As these networks pledge to boost speed, capacity, latency, and connectivity, the challenges of effective integration in densely urbanized environments become evident. Hence, addressing these challenges, from efficient spectrum management to enhanced coverage and service quality, demands an innovative approach. Recently, various studies on reconfigurable intelligent surfaces (RIS) design have been performed by both industry and academia. For instance, field measurement campaigns on commercial mobile networks, system-level simulations, designs, fabrications, and individual experimental demonstrations spanning sub-6GHz, millimeter wave (mmWave), and sub-THz bands. Specifically, in [1], the authors performed a field measurement campaign deploying manufactured/tuned RIS at 2.6 GHz to improve coverage in dense areas. Likewise, in [2], a prototype RIS was developed at the level of macro-urban environments to enhance the performance of the underlying system in terms of signal-to-interference-plus-noise ratio (SINR), physical resource blocks (PRBs) usage, and coverage by using $3.5/28$ GHz bands. In [3], a RIS design was carried out by employing mmWave frequencies (e.g., $27.5$ to $29.5$ GHz) with real-world applications in 5G and B5G networks, showcasing the versatility of RIS applications [3]. Simulation results and experimental campaigns were carried out in [4] at frequencies of $2.6$, $4.9$, and 26 GHz with $1$ and 2-bit arrays to explore how to improve spectral efficiency performance. The evolution of RIS opens the door to simultaneously transmitting and reflecting intelligent omnisurfaces (STAR-IOS) achieving benefits like decoupling of transmission/reception and improved coverage patterns [5]. Additionally, designs, Ray Tracing modeling, and simulations in the FR2 band implemented in a commercial network showed improvements in coverage, quality, and capacity network [6]. Analyses in sub-6 GHz bands [7] and practical demonstrations in various use cases for sixth-generation (6G) mobile networks [8] underscore the expanding scope of RIS applications. With a lack of applied research in planning/optimization for mobile cellular networks using RIS tested in Latin America, this work aims to fill this gap by applying the concepts of manufactured/tuned RIS in the $3.5/28$ GHz bands to design/deploy a commercial 5G/B5G network for Quito, Ecuador. In particular, this paper presents an advanced exploration of optimizing 5G/B5G networks in dense urban areas. For this purpose, a key innovation is the integration of RIS, strategically positioned to enhance coverage and spectral efficiency at critical frequency bands, i.e., $3.5/28$ GHz. In the proposed network design, the synchronized transmission with advanced reconfigurable surfaces (STARS) is employed as an innovative strategy that addresses urban density challenges and promises revolutionary improvements in signal quality and spectral efficiency. This strategy contributes significantly to the transition from 5G/B5G to the envisioned landscape of 6G networks. The refined methodology aims to elevate connectivity and network capacity in complex urban environments, as the Quito case study exemplifies. Finally, useful insights on design criteria when relying on STAR-RIS are also provided.

The organization of this paper is as follows: Section II describes some typical dead zones with coverage gaps in a commercial 5G network and the current methodologies for planning/optimizing a 5G/B5G network. Section III proposes the methodology for configuring a commercial 5G/B5G network assisted by STAR-RIS in the 3.5/28 GHz bands. Section IV applies the design to the case study of the city of Quito. Next, the results and discussions are presented in Section V, and finally, concluding remarks are provided in Section VI.

## II. Dead Zones And Coverage Overshadow Areas In 5G/B5G Networks

### A. Urban/Dense Urban Challenges in 5G/B5G Networks

In dead zones of commercial mobile networks with weak coverage, holes are inevitable due to physical propagation limitations, such as the limited down-tilt angle of base station (BS) antennas and high-penetration loss caused by buildings. Specific dead zones include the under- tower shadow zone, characterized by low signal strength below the BS, and indoor shadow zones with signal attenuation through obstacles like building walls. Also, outdoor building shadow zones near tall structures (e.g., towers, monopoles) hinder direct wave propagation [1]; these cases are viewed in Fig. 1. Commonly, in commercial 5G/B5G networks, researchers are focusing on mobility and coverage performance. In areas with heavy traffic and dense buildings/radio towers, vehicles and pedestrians move slowly, leading to more time spent in these zones/areas. These holes and shadow areas occur on highways, avenues, sidewalks, pavements, and city streets. A typical solution would be to install more Next Generation Nodes B (gNodeBs); however, unmanned aerial vehicles (UAVs) can serve these dead zones and shadow areas, for disaster management scenarios. UAVs have efficiently improved the reference signal received power (RSRP) levels in real-world cases [9]. Deploying small BSs at the edge of macro cells or in traffic hot spots may not adequately cover vehicle-to-network communication, especially at mmWave frequencies [10]. Street coverage often has dead zones and empty spaces, particularly at intersections, which are much more noticeable when operating with mmWave bands. While deploying ultra-dense small BSs improves coverage at 28 GHz, it still needs to guarantee consistently high data rates due to remaining coverage gaps. In this context, notable improvements concerning the signal-to-interference-plus-noise ratio (SINR) and coverage have been achieved with the densification of small cells, primarily in sub-6GHz and mmWave bands [10]. In order to address these coverage issues, our work explores an RIS-assisted commercial system designed to enhance coverage/throughput in dead zones. In subsequent sections, we provide detailed information on the planning/optimizing RIS-assisted commercial techniques. Then, we perform simulation plots to explore the underlying performance in real networks.

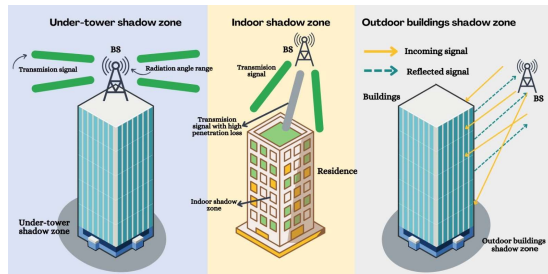

Fig. 1. Typical dead zones: under-tower, indoor and outdoor buildings shadow zones [10].

### B. Efficient Coverage Improvement using RIS Solutions

The proposed RIS-assisted 6G network consists of i) macrocells/microcells, which are equipped with gNodeBs, ii) the user equipment (UE), and iii) the RIS technology. While the UE and the gNodeB are inherent to the mobile network, the RIS is introduced to enhance communication efficiency between the gNodeB and the UE, primarily in scenarios with obstructed signals due to buildings and obstacles, including shadows/hole zones. According to [1], the standard architecture of the RIS consists of a thin two-dimensional structure with an array system composed of multiple unit cells the size of a sub-wavelength. Each RIS unit cell can be controlled electronically using a field-programmable gate array (FPGA) and a control unit with as many outputs as bits are required per unit cell. The RIS unit cell can be controlled using varactor diodes, pin diodes, liquid crystal, and liquid metal, to mention a few options [11]. This way, the signal propagation environment between the gNodeB and the UE, is controlled by gNodeB with RIS. Additionally, to further improve efficiency in covering shadow areas, we propose using STAR-RIS, which is essential in highly obstructed environments with high user density and when implementing several simultaneous RIS in enclosed environments. In recent years, prototypes of RIS have been developed, and their performance has been enhanced when used in commercial 5G networks. These RIS setups can address signaling, synchronization, RIS size, bit configuration, reflection and diffraction degrees, scattering, working frequencies, number of elements, energy safety, and acceptable gains [1]–[7]. In the state-of-the-art, a large body of RIS works is dedicated to channel modeling, system- level simulations, individual experimental tests, channel training, and network coexistence. However, due to the lack of research in Latin America for the deployment of STAR-RIS in commercial 5G/B5G networks, we propose a methodology for the practical use of standardized RIS to efficiently plan/optimize 5G radio network focusing on dead zone cases due to coverage shadows in Quito, Ecuador at the 3.5/28 GHz bands. According to the primary research and studies carried out on RIS (e.g., designs, channel/propagation modeling, deployments, field trials, simulations), we summarize in Table I the system configuration to be taken into account to apply, at a macro level, the use of STAR-RIS in the proposed system. Such a setup provides efficiency to the radio planning/optimization of 5G/6G networks, starting from RIS to STAR-RIS on Line-of-Sight (LOS) and Non-Line-of-Sight (NLOS) scenarios, as illustrated in Fig. 2 and Fig. 3. Similarly, the hardware devices of the selected RISs for the proposed 5G network design/optimization are presented in Fig. 4, where the characteristics of the RIS for 3.5/28 GHz and their applicability in the B5G mobile network are observed.

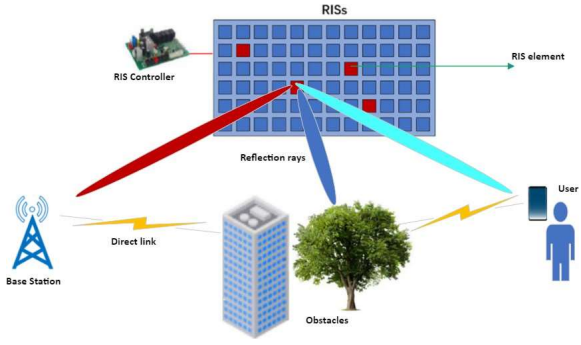

Fig. 2. A typical architecture of a RIS with MIMO system [8].

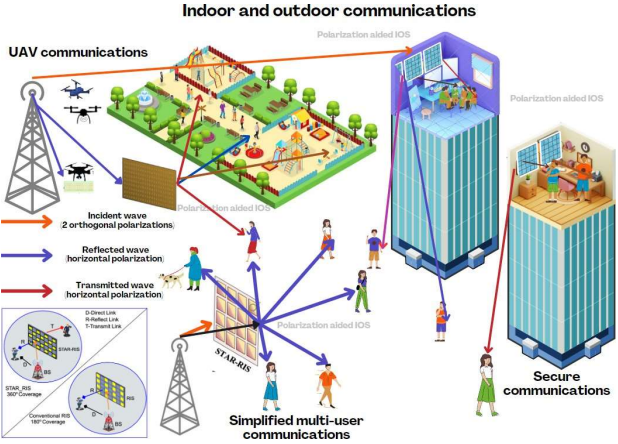

Fig. 3. Scenarios of application of the STAR-RIS with IOS in wireless communications [5] [12].

## C. Classic Methodology for 5G Planning/Optimization

Based on the consolidated practices of operators and vendors in real 5G/B5G commercial networks and recent studies, the classic methodology for radio access network (RAN) planning/optimization can be summarized according to the process indicated in [13], which begins with the collection of all necessary operator information. Then, we proceed to the network planning processes by calculating the link budget, cell radius, necessary coverage area, and minimum number of nodes for coverage. As the next step, we perform capacity planning by determining peak/average throughput, number of users per service, and minimum number of users for network capacity compliance. Finally, coverage, quality, and capacity simulations are conducted, considering the correct geographical placement of sites, with corresponding configurations of physical radio frequency parameters, logical configurations, and basic/advanced functionalities. Therefore, the optimization process is carried out by improving the results obtained in coverage/capacity design through various techniques ranging from deploying a new macro/micro site to using indoor solutions. This process is iterative and feedback-driven with results.

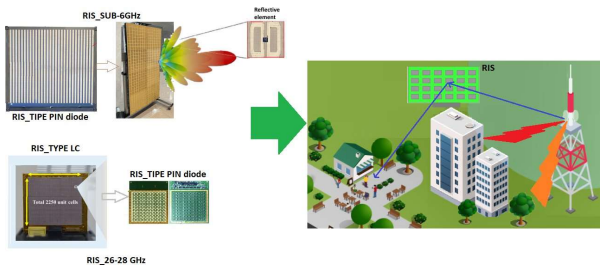

Fig. 4. Selected RIS for 3.5 GHz/28 GHz [1], [3], [4], [6], [7].

Table I
MAIN SETTING PARAMETERS FOR RIS/STAR-RIS [1]–[7]

| Item N° | RIS/STAR-RIS Parameters Name | 3.5 GHz | 28 GHz |
|---|---|---|---|
| 1 | Central Frequency (GHz) | 3.5 | 28 |
| 2 | Carrier Bandwidth (MHz) | 100 | 200 |
| 3 | number of total antenna elements | 2430 | 2250 |
| 4 | Working/Control voltage (V) | 10.5 | 5 – 10 |
| 5 | Consumption power (W) | 3 – 6 | 3 – 6 |
| 6 | RIS width/square shape mxm | 3.8×3.8 | 0.67×0.67 |
| 7 | number of bits/phase resolution | 1 | 2 |
| 8 | number of BS | 8 | 140 |
| 9 | BS height (m) | 25 | 25 |
| 10 | BS Tx Power (dBm) | 18 | 18 |
| 11 | UE Noise figure (dB) | 5 | 5 |
| 12 | RIS reflection losses (dB) | 0 – 1 | 2.68 – 3.05 |
| 13 | Channel models | 3GPP | 3GPP |
| 14 | Cell edge target SINR (dB) | 10 | 10 |
| 15 | Phase diff. ON-OFF(°) | 160 | 160 – 180 |
| 16 | RIS Gain (dB) | 15 | 26 – 30.8 |
| 17 | BS-RIS distance(m) | 29.75 | 20 |
| 18 | RIS-UE distance(m) | 10 – 20 | 10 |
| 19 | RIS Type | PIN-d | PIN-d,LC |
| 20 | Incidence angle(°) | 15 – 45 | 10 – 60 |

## III. METHODOLOGY FOR RAN PLANNING/OPTIMIZATION USING RIS/STARS

According to [13], our scheme of the 5G model proposes a new and improved methodology for the planning and optimization process of a 5G cellular network, aiming at establishing an efficient advanced 5G network and laying a solid foundation for designing and deploying a pre-evolved 6G network. In Fig. 5, we show the diagram to be applied to a case study in Quito city, starting from a 5G network designed and simulated for a dense urban central area in the 3.5/28 GHz bands. Our main goal is to optimize the proposed 5G system and evolve it into a potential B5G network for the new challenges posed by promising key technologies that the upcoming 6G network brings.

### A. Methodology Description and Key Considerations

In this paper's proposed methodology for efficient planning/optimization of 5G/B5G radio networks, we have added specific procedures for STAR-RIS. Thus, the following processes have been included: selection of the adjusted propagation model, calculation of path loss/cell radius, consideration of gains/losses, and calculation from propagation simulations of coverage areas/zones. It is worth mentioning that the geographic locations of RISs in coverage shadow areas have also been considered, along with the calculation of the number of RISs required to achieve the desired coverage probability percentage (e.g., 95%). The incidence/reflection angles, heights, phase differences, and transmission (Tx) powers are also considered. Finally, the selectivity of RIS/STARs usage, potential relocations thereof, and other modern coverage/capacity optimization techniques for 5G/B5G have been accounted for. Consequently, the final simulation results incorporate the joint propagation of RIS/STARs added to the initially 5G planned sites.

### B. RIS Propagation Channel Modeling

In Fig. 6, we illustrate a wireless communication system in which RIS helps communicate between the BS and the UE. To do this, RIS introduces scattering gain depending on its size and material. Standardization of the type of material is necessary to encourage the use of RIS. Therefore, modifications to the 3rd Generation Partnership Project (3GPP) channel generation procedure have been made recently, including path loss, arrival and departure angles, and reflection function [13]. The path loss calculation is adjusted depending on whether RIS is broadcasting (near-field) or beamforming (far-field). On the other hand, studies such as [14] indicated that RIS-assisted channels typically involve multi-segment channels, altering the path loss representation. The relationship between distances from BS to RIS (dtx) and from RIS to UE (drx) is crucial for path loss modeling. The RIS-assisted channel generally contains multi-segment channels. The introduction of RIS provides a new dimension of variations for signal channels. Consequently, a single distance between BS and UE can no longer represent the channel path loss. In a "BS-RIS-UE" channel, the received signal power is related to dtx and drx. The relationship between dtx and drx is the key to path loss modeling for RIS-assisted channels [14]. Clusters of a channel between BS and RIS and a channel between BS and UE can be independently modeled. Similarly, the angle of arrival and the angle of departure can be calculated by reflecting their locations rather than randomly computing them [15]. Path loss modeling varies depending on far-field, beamforming, and near-field scattering. In the case of near-field scattering, additive fading is reflected instead of multiplicative fading in path loss. Consequently, the RIS channel model can be generated by reflecting the above factors in the 3GPP channel model. Thus, dtx and drx represent RIS-assisted channel path loss in additive and multiplicative relationships. Furthermore, recent studies mainly consider dtx and drx to be more multiplicative relationships. Thus, in the case of far-field or RIS beamforming, the additive fading in [15] is replaced with multiplicative fading according to the mathematical expression given by

$$PL_{Tot1} = 10\log_{10}(10^{\frac{PL_{B,R}}{10}} \times 10^{\frac{PL_{R,U}}{10}})[dB], \quad (1)$$

where:

$PL_{B,R}$ is the pathloss value between BS and RIS.

$PL_{R,U}$ is the pathloss value between RIS and UE.

According to [14], a valid formulation for obtaining an RIS-assisted channel path loss model, taking into account the difference between the BS-RIS channel and the RIS-UE channel, having a better fitting effect than the multiplicative expression, and is given by:

$$PL_{Tot2} = 24\log_{10}(dtx) + 19.2\log_{10}(drx) + 63.22[dB], \quad (2)$$

where:

dtx is the distance between BS and RIS (28GHz).

drx is the distance between RIS and UE (28GHz).

In (2), we cannot directly explain its physical correctness due to the uncertainty of how different channels affect the model. Therefore, we chose (2) as the optimal model among these three [14] for the mmWave band and (1) for the sub-6GHz band.

## IV. ENHANCED PLANNING/OPTIMIZATION FOR 5G/B5G: CASE STUDY QUITO CITY

### A. Practical Application in Case study of Quito city

As an initial step and available information, we start from the same target area in the dense urban area of Quito, according to [13]. The initially designed 5G network is characterized by being a Non-Stand Alone (NSA) architecture network, option 3X sized to provide services in the enhanced Mobile BroadBand use case, with functionalities from 3GPP releases 15/16. This 5G planned network has 8 gnodesB in 3.5 GHz and 140 gnodesB in 28 GHz. By performing planning/optimization efforts using RIS/STARs, we take a look at what it will be like for B5G and 6G networks to provide services in other use cases, such as ultra-reliable low-latency/massive-IoT utilizing releases 17/18, with performance in coverage/quality/capacity. Additionally, through other communication techniques such as Holographic massive Multiple-Input-Multiple-Output (mMIMO), fluid antenna systems, and others, we will equip the designed 5G/B5G network with key features to evolve into 6G. Utilizing the study conducted in [13], and after considering the simulation plots obtained from coverage/quality/capacity in the Atoll tool for a 5G network previously designed in the frequencies of 3.5/28 GHz, we proceeded to carry out all the steps and calculation processes by the proposed methodology introduced in Fig. 5. Thus, based on the descriptions provided in sections III A-B, the results shown in Table II have been determined for the chosen RIS/STARs in our design/optimization process. As a next step, we determine the dead zones and shadow areas, starting from the input information of the initial coverage plot (RSRP) of the reference 5G network design according to [13], as shown in Fig. 7. Once the main calculations regarding the RIS/STAR are determined, as well as the potential shadow areas/coverage gaps in our initial 5G design: The placement of the RISs/STARs is developed, geographically locating them in the Atoll tool according to the data in Table II and in accordance with the coverage requirement indicated in Fig. 7. In this way, we perform simulations of the respective plots with/without RIS/STARs for coverage, quality, and capacity of the designed/optimized 5G network. Where the all calculations from Fig. 5 were performed using the Atoll and Excel tools.

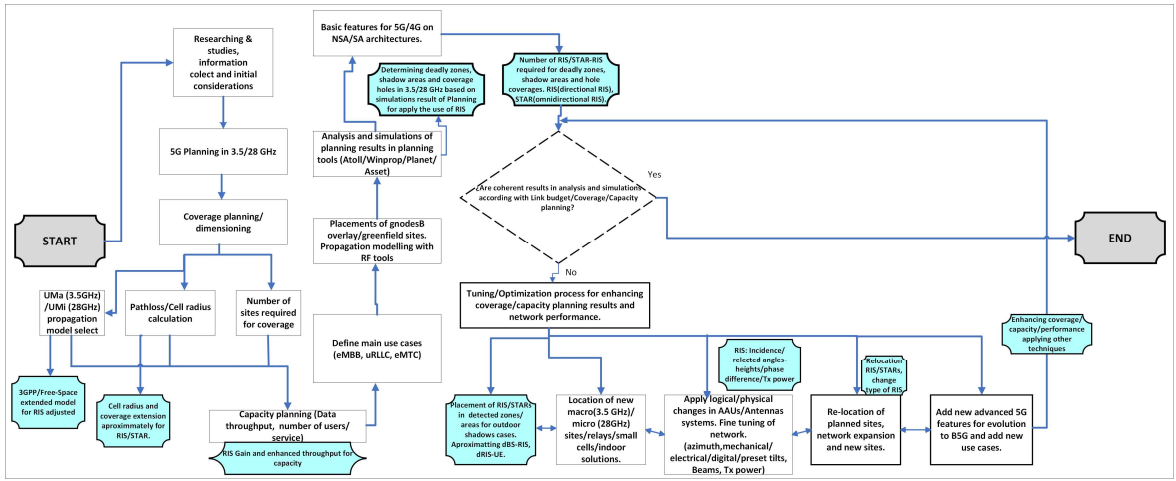

Fig. 5. Proposed methodology for RAN Planning/Optimization with RIS [13].

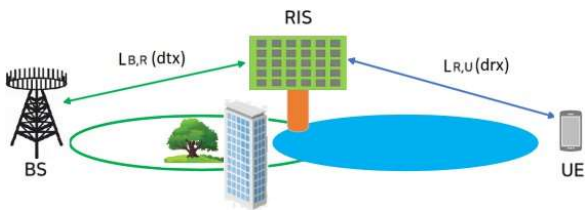

Fig. 6. Propagation scenario with RIS, in based to [15].

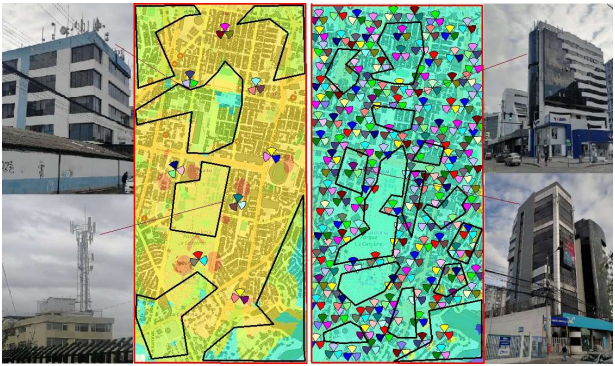

Fig. 7. Dead/shadow zones 3.5 GHz (left) and 28 GHz (right).

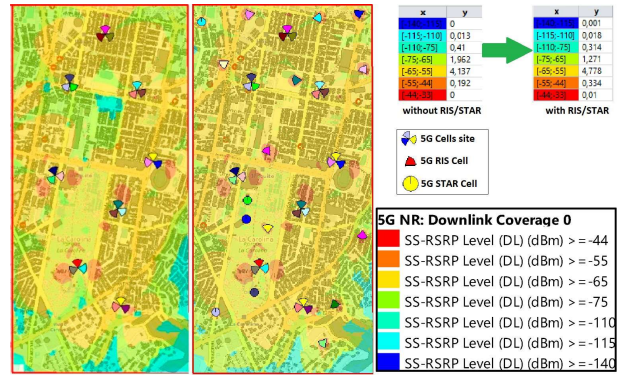

Fig. 8. RSRP plot 3.5 GHz, without RISs (left) and with RISs (right).

Table II
MAIN CALCULATIONS FOR RIS/STAR-RIS (PROPOSED METHODOLOGY)

| Item N° | RIS/STAR-RIS 5G Planning/Optimization calculations | | |
|---|---|---|---|
| | *Process* | *3.5 GHz* | *28 GHz* |
| 1 | 5G Channel modeling | 3GPP ext | Corridor ext |
| 2 | Location of deadly zones | Fig.7(left) | Fig.7(right) |
| 3 | Approx. cell radius (m) | 300 | 70 |
| 4 | Coverage area needed (Km²) | 2.48 | 1.86 |
| 5 | Min. number of RISs needed | 11 | 142 |
| 6 | Gain/approx. improvement (dB) | 10 | 10 |
| 7 | Placement of RISs | Fig.8(right) | Fig.10(right) |
| 8 | Incid./reflect. angles(°) | 45/160 | 10/160 |
| 9 | $d_{BIS-RIS} / d_{RIS-UE}$ (m) | 100-200 | 50-75 |
| 10 | Heights(m)/Tx Power(dBm) | 5/18 | 5/21 |

## V. RESULTS AND DISCUSSIONS

In Fig. 8, it can be observed that our coverage plot in 3.5 GHz with the design/optimization of the 5G network using RIS increases the good and excellent levels of RSRP by 30% to 40% in the target area (6.72Km²), compared to the initial design without using RIS. On the other hand, in Fig. 9, we illustrate the Throughput for 3.5GHz; here, it can be verified that there is a 15% improvement in capacity coverage in the target area using RIS. Also, note that the average throughputs are above 100 Mbps per UE. In Fig 10, it is observed that for the 28 GHz band, there is an increase in high and medium RSRP levels ranging from 50 to 200% and a decrease in low and very low levels. Similarly, in the metric curves of Fig. 11, it is noticed that when using RIS, there is an improvement between 80-400% in very high, high, and medium SINR levels with a decrease in low-quality SINR levels in the 28GHz band. Finally, in Fig. 12, the throughput plots at 28GHz show that a 15% improvement in throughput coverage is achieved using RIS.

## VI. CONCLUSIONS AND FUTURE DIRECTIONS

This work demonstrated that by incorporating RIS in the 5G network for coverage/quality/throughput optimization in the 28GHz band, better results were obtained without compromising quality levels than those obtained with STAR-

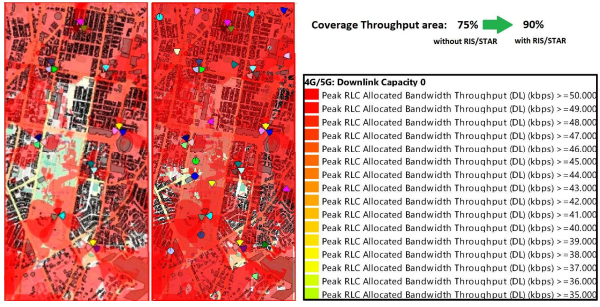

Fig. 9. Throughput plot 3.5 GHz, without RISs (left)/with RISs (right).

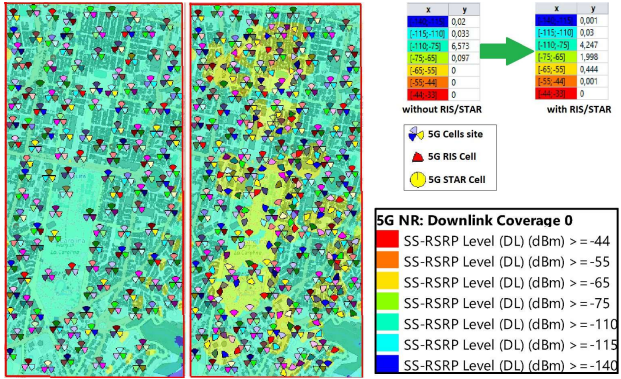

Fig. 10. RSRP plot 28 GHz, without RISs (left) and with RISs (right).

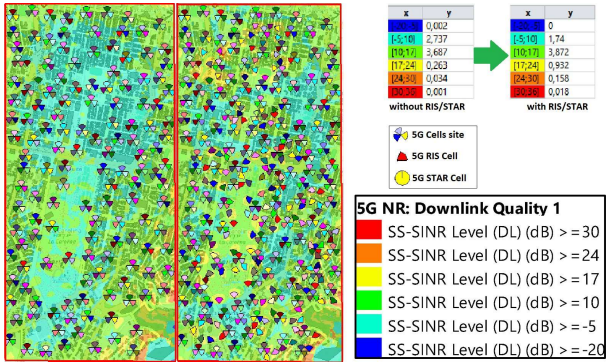

Fig. 11. SINR plot 28 GHz, without RISs (left) and with RISs (right).

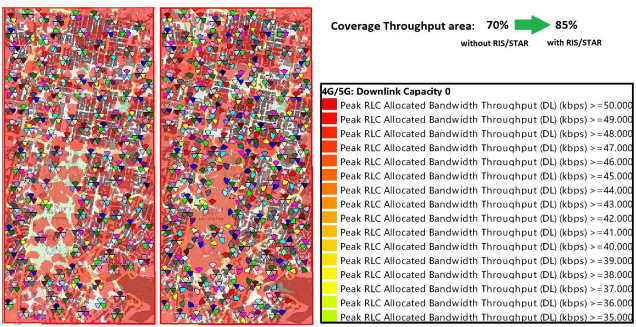

Fig. 12. Throughput plot 28 GHz, without RISs (left) and with RISs (right).

RIS in the 3.5GHz band. Overall, it was revealed that the use of STAR-RIS in planning processes improved the performance of coverage and throughput levels, mainly for shadow areas where UEs are affected under buildings and beneath communication towers in the external zones of the target area. Also, it can be highlighted that using STAR-RIS for the proposed planning and optimization procedures of 5G/B5G networks contributed to providing efficiency before, during, and after commercial deployment in real-life scenarios for mobile operators. An interesting future work could be to perform planning/optimizations considering interference environments in massive deployments of RIS, holographic Multiple-Input Multiple-Output (MIMO), and UAV-mounted RIS. Also, the critical implications of multi-user environments for dense urban scenarios could be studied and analyzed.